\documentclass[aps,prd,twocolumn,superscriptaddress,showpacs,letterpaper]{revtex4}

\usepackage{natbib}
\usepackage{graphicx} 
\usepackage{bbm}
\usepackage{hyperref}
\usepackage{mycaption}

\newcommand{\tbar}{\ensuremath \overline{t}}
\newcommand{\ttbar}{\ensuremath {t\tbar}}
\newcommand{\pbar}{\ensuremath \overline{p}}
\newcommand{\ppbar}{\ensuremath {p\pbar}}

\newcommand{\met}{{\ensuremath \not\!\!E_{\mathrm{T}}}}
\newcommand{\gev}{\ensuremath \mathrm{GeV}}

\newcommand{\invfb}{\ensuremath \mathrm{fb^{-1}}}

\newcommand{\pb}{\ensuremath \mathrm{pb}}

\newcommand{\ra}{\ensuremath\rightarrow}
\newcommand{\goesto}{\ensuremath\!\ra\!}

\begin{document}

\begin{minipage}{\textwidth}
  \begin{flushright}
    FERMILAB-PUB-11-350-E\\
    \vspace*{0.25cm}
  \end{flushright}
\end{minipage}

\vfill

\affiliation{Institute of Physics, Academia Sinica, Taipei, Taiwan 11529, Republic of China}
\affiliation{Argonne National Laboratory, Argonne, Illinois 60439, USA}
\affiliation{University of Athens, 157 71 Athens, Greece}
\affiliation{Institut de Fisica d'Altes Energies, ICREA, Universitat Autonoma de Barcelona, E-08193, Bellaterra (Barcelona), Spain}
\affiliation{Baylor University, Waco, Texas 76798, USA}
\affiliation{Istituto Nazionale di Fisica Nucleare Bologna, $^{cc}$University of Bologna, I-40127 Bologna, Italy}
\affiliation{University of California, Davis, Davis, California 95616, USA}
\affiliation{University of California, Los Angeles, Los Angeles, California 90024, USA}
\affiliation{Instituto de Fisica de Cantabria, CSIC-University of Cantabria, 39005 Santander, Spain}
\affiliation{Carnegie Mellon University, Pittsburgh, Pennsylvania 15213, USA}
\affiliation{Enrico Fermi Institute, University of Chicago, Chicago, Illinois 60637, USA}
\affiliation{Comenius University, 842 48 Bratislava, Slovakia; Institute of Experimental Physics, 040 01 Kosice, Slovakia}
\affiliation{Joint Institute for Nuclear Research, RU-141980 Dubna, Russia}
\affiliation{Duke University, Durham, North Carolina 27708, USA}
\affiliation{Fermi National Accelerator Laboratory, Batavia, Illinois 60510, USA}
\affiliation{University of Florida, Gainesville, Florida 32611, USA}
\affiliation{Laboratori Nazionali di Frascati, Istituto Nazionale di Fisica Nucleare, I-00044 Frascati, Italy}
\affiliation{University of Geneva, CH-1211 Geneva 4, Switzerland}
\affiliation{Glasgow University, Glasgow G12 8QQ, United Kingdom}
\affiliation{Harvard University, Cambridge, Massachusetts 02138, USA}
\affiliation{Division of High Energy Physics, Department of Physics, University of Helsinki and Helsinki Institute of Physics, FIN-00014, Helsinki, Finland}
\affiliation{University of Illinois, Urbana, Illinois 61801, USA}
\affiliation{The Johns Hopkins University, Baltimore, Maryland 21218, USA}
\affiliation{Institut f\"{u}r Experimentelle Kernphysik, Karlsruhe Institute of Technology, D-76131 Karlsruhe, Germany}
\affiliation{Center for High Energy Physics: Kyungpook National University, Daegu 702-701, Korea; Seoul National University, Seoul 151-742, Korea; Sungkyunkwan University, Suwon 440-746, Korea; Korea Institute of Science and Technology Information, Daejeon 305-806, Korea; Chonnam National University, Gwangju 500-757, Korea; Chonbuk National University, Jeonju 561-756, Korea}
\affiliation{Ernest Orlando Lawrence Berkeley National Laboratory, Berkeley, California 94720, USA}
\affiliation{University of Liverpool, Liverpool L69 7ZE, United Kingdom}
\affiliation{University College London, London WC1E 6BT, United Kingdom}
\affiliation{Centro de Investigaciones Energeticas Medioambientales y Tecnologicas, E-28040 Madrid, Spain}
\affiliation{Massachusetts Institute of Technology, Cambridge, Massachusetts 02139, USA}
\affiliation{Institute of Particle Physics: McGill University, Montr\'{e}al, Qu\'{e}bec, Canada H3A~2T8; Simon Fraser University, Burnaby, British Columbia, Canada V5A~1S6; University of Toronto, Toronto, Ontario, Canada M5S~1A7; and TRIUMF, Vancouver, British Columbia, Canada V6T~2A3}
\affiliation{University of Michigan, Ann Arbor, Michigan 48109, USA}
\affiliation{Michigan State University, East Lansing, Michigan 48824, USA}
\affiliation{Institution for Theoretical and Experimental Physics, ITEP, Moscow 117259, Russia}
\affiliation{University of New Mexico, Albuquerque, New Mexico 87131, USA}
\affiliation{The Ohio State University, Columbus, Ohio 43210, USA}
\affiliation{Okayama University, Okayama 700-8530, Japan}
\affiliation{Osaka City University, Osaka 588, Japan}
\affiliation{University of Oxford, Oxford OX1 3RH, United Kingdom}
\affiliation{Istituto Nazionale di Fisica Nucleare, Sezione di Padova-Trento, $^{dd}$University of Padova, I-35131 Padova, Italy}
\affiliation{University of Pennsylvania, Philadelphia, Pennsylvania 19104, USA}
\affiliation{Istituto Nazionale di Fisica Nucleare Pisa, $^{ee}$University of Pisa, $^{ff}$University of Siena and $^{gg}$Scuola Normale Superiore, I-56127 Pisa, Italy}
\affiliation{University of Pittsburgh, Pittsburgh, Pennsylvania 15260, USA}
\affiliation{Purdue University, West Lafayette, Indiana 47907, USA}
\affiliation{University of Rochester, Rochester, New York 14627, USA}
\affiliation{The Rockefeller University, New York, New York 10065, USA}
\affiliation{Istituto Nazionale di Fisica Nucleare, Sezione di Roma 1, $^{hh}$Sapienza Universit\`{a} di Roma, I-00185 Roma, Italy}
\affiliation{Rutgers University, Piscataway, New Jersey 08855, USA}
\affiliation{Texas A\&M University, College Station, Texas 77843, USA}
\affiliation{Istituto Nazionale di Fisica Nucleare Trieste/Udine, I-34100 Trieste, $^{ii}$University of Udine, I-33100 Udine, Italy}
\affiliation{University of Tsukuba, Tsukuba, Ibaraki 305, Japan}
\affiliation{Tufts University, Medford, Massachusetts 02155, USA}
\affiliation{University of Virginia, Charlottesville, Virginia 22906, USA}
\affiliation{Waseda University, Tokyo 169, Japan}
\affiliation{Wayne State University, Detroit, Michigan 48201, USA}
\affiliation{University of Wisconsin, Madison, Wisconsin 53706, USA}
\affiliation{Yale University, New Haven, Connecticut 06520, USA}

\author{T.~Aaltonen}
\affiliation{Division of High Energy Physics, Department of Physics, University of Helsinki and Helsinki Institute of Physics, FIN-00014, Helsinki, Finland}
\author{B.~\'{A}lvarez~Gonz\'{a}lez$^x$}
\affiliation{Instituto de Fisica de Cantabria, CSIC-University of Cantabria, 39005 Santander, Spain}
\author{S.~Amerio}
\affiliation{Istituto Nazionale di Fisica Nucleare, Sezione di Padova-Trento, $^{dd}$University of Padova, I-35131 Padova, Italy}
\author{D.~Amidei}
\affiliation{University of Michigan, Ann Arbor, Michigan 48109, USA}
\author{A.~Anastassov$^v$}
\affiliation{Fermi National Accelerator Laboratory, Batavia, Illinois 60510, USA}
\author{A.~Annovi}
\affiliation{Laboratori Nazionali di Frascati, Istituto Nazionale di Fisica Nucleare, I-00044 Frascati, Italy}
\author{J.~Antos}
\affiliation{Comenius University, 842 48 Bratislava, Slovakia; Institute of Experimental Physics, 040 01 Kosice, Slovakia}
\author{G.~Apollinari}
\affiliation{Fermi National Accelerator Laboratory, Batavia, Illinois 60510, USA}
\author{J.A.~Appel}
\affiliation{Fermi National Accelerator Laboratory, Batavia, Illinois 60510, USA}
\author{T.~Arisawa}
\affiliation{Waseda University, Tokyo 169, Japan}
% ...

\collaboration{CDF Collaboration\footnote{With visitors from
$^a$Istituto Nazionale di Fisica Nucleare, Sezione di Cagliari, 09042 Monserrato (Cagliari), Italy,
$^b$University of CA Irvine, Irvine, CA 92697, USA,
$^c$University of CA Santa Barbara, Santa Barbara, CA 93106, USA,
$^d$University of CA Santa Cruz, Santa Cruz, CA 95064, USA,
$^e$CERN, CH-1211 Geneva, Switzerland,
$^f$Cornell University, Ithaca, NY 14853, USA,
$^g$University of Cyprus, Nicosia CY-1678, Cyprus,
$^h$Office of Science, U.S. Department of Energy, Washington, DC 20585, USA,
$^i$University College Dublin, Dublin 4, Ireland,
$^j$ETH, 8092 Zurich, Switzerland,
$^k$University of Fukui, Fukui City, Fukui Prefecture, Japan 910-0017,
$^l$Universidad Iberoamericana, Mexico D.F., Mexico,
$^m$University of Iowa, Iowa City, IA 52242, USA,
$^n$Kinki University, Higashi-Osaka City, Japan 577-8502,
$^o$Kansas State University, Manhattan, KS 66506, USA,
$^p$University of Manchester, Manchester M13 9PL, United Kingdom,
$^q$Queen Mary, University of London, London, E1 4NS, United Kingdom,
$^r$University of Melbourne, Victoria 3010, Australia,
$^s$Muons, Inc., Batavia, IL 60510, USA,
$^t$Nagasaki Institute of Applied Science, Nagasaki, Japan,
$^u$National Research Nuclear University, Moscow, Russia,
$^v$Northwestern University, Evanston, IL 60208, USA,
$^w$University of Notre Dame, Notre Dame, IN 46556, USA,
$^x$Universidad de Oviedo, E-33007 Oviedo, Spain,
$^y$CNRS-IN2P3, Paris, F-75252 France,
$^z$Texas Tech University, Lubbock, TX 79609, USA,
$^{aa}$Universidad Tecnica Federico Santa Maria, 110v Valparaiso, Chile,
$^{bb}$Yarmouk University, Irbid 211-63, Jordan,
$^{jj}$On leave from J.~Stefan Institute, Ljubljana, Slovenia
}}
\noaffiliation

\begin{abstract}
We search for resonant production of $t\bar{t}$ pairs in $4.8\ \mathrm{fb^{-1}}$
integrated luminosity of $p\bar{p}$ collision data at $\sqrt{s}=1.96\
\mathrm{TeV}$ in the lepton+jets decay channel, where one top quark decays
leptonically and the other hadronically. A matrix element reconstruction
technique is used; for each event a probability density function (pdf) of the
$t\bar{t}$ candidate invariant mass is sampled. These pdfs are used to construct
a likelihood function, whereby the cross section for resonant $t\overline{t}$
production is estimated, given a hypothetical resonance mass and width. The data
indicate no evidence of resonant production of $t\bar{t}$ pairs. A benchmark
model of leptophobic $Z'\!\ra\! t\bar{t}$ is excluded with $m_{Z'} < 900
\ \mathrm{GeV}/c^2$ at 95\% confidence level. 
\end{abstract}

\title{A search for resonant production of $t\bar{t}$ pairs in $4.8\
  \rm{fb}^{-1}$ of integrated luminosity of $p\bar{p}$ collisions at
  $\sqrt{s}=1.96\ \rm{TeV}$} 

\hypersetup{colorlinks=true,backref=page,breaklinks=false,%
  linkcolor=blue,urlcolor=blue,citecolor=blue,%
  pdftitle={A search for resonant production of ttbar pairs in 4.8 fb^{-1} of
    integrated luminosity of ppbar collisions at sqrt(s)=1.96 TeV},
  pdfauthor=the CDF Collaboration}


\maketitle

Many theories of physics beyond the standard model (SM) predict additional
vector bosons (e.g. $Z'$). These include (but are not limited to) extended
gauge theories (e.g. $\mathrm{SO}(10)$), Kaluza--Klein states of the gluon
or of $Z$ bosons, axigluons, and topcolor~\cite{Weinberg, Dimo_georgi,
Hill:1993hs, Arkani-Hamed_1, Randall_Sundrum, topcolor_Zprime, Chivukula,
Arkani-Hamed_0, Fitzpatrick:2007qr, Frederix:2007gi}. Previous analyses of
Tevatron data have excluded such narrow resonances with masses less than $725 \
\gev\!/c^2$~\cite{10.1103/PhysRevLett.92.221801, PhysRevD.77.051102,
  PhysRevLett.100.231801, j.physletb.2008.08.027}. 

In this Letter we describe a search for narrow resonant states decaying to
top--antitop pairs in proton--antiproton collisions using the CDF II detector 
at the Tevatron at Fermilab. We discuss the experimental signature of resonant
$\ttbar$ production and how we are to distinguish it from standard model
$\ttbar$ production. We describe the reconstruction technique used in this
analysis and the statistical tests we perform in examining the data for any sign
of the hypothetical resonant production, and we describe the systematic
uncertainties related to this analysis. Observing no evidence for resonant
production of $\ttbar$ pairs, we set upper limits on the cross section times
branching ratio for resonant $\ttbar$ production. 
 
At the Tevatron, the $\ttbar$ pair--production cross section has been measured
with great precision: $\sigma_{\ppbar\goesto\ttbar} = 7.70\pm 0.52\
\mathrm{pb}$~\cite{ttbar_xsection}. 
However, this degree of precision leaves open the possibility that non--SM
physics gives rise to a fraction of the total $\ttbar$ production. 

We search for a heavy vector boson decaying to $\ttbar$
in the final state where one top quark decays semileptonically 
($t\goesto\ell\nu b$) and the other hadronically ($t\goesto
q\overline{q}'b$)~\cite{lepton_jets_note} by examining the
$\ttbar$ invariant mass spectrum of candidate events, where the event kinematics
have been reconstructed by applying the SM QCD ``Matrix Element'' for $\ttbar$
production and decay~\cite{PhysRevLett.100.231801}.  
The observed spectrum is then compared to templates --
models of signal (i.e., $Z'\goesto \ttbar$) and background processes 
(e.g. SM $\ttbar$, $W$+jets, $WW/WZ$) in an unbinned maximum--likelihood
fit. By fitting the data to these models, we extract upper limits on
the $Z'\goesto\ttbar$ cross section times branching ratio. We consider scenarios
where the $Z'$ width is 1.2\% of the pole mass (this has been the benchmark
scenario for narrow--resonance searches in $\ttbar$ enriched
samples~\cite{topcolor_Zprime}).

The CDF detector is a general purpose, azimuthally and forward-backward
symmetric multipurpose collider detector. A detailed description can be found in
Ref.~\cite{CDFII_detector}; here we summarize details of detector components
important to this analysis. The transverse momenta ($p_\mathrm{T}$) and track
parameters of charged particles are measured by an eight-layer silicon strip
detector~\cite{L00, SVXII, ISL} and a 96-layer drift chamber (COT)~\cite{COT},
both within a 1.4~T magnetic field. The COT provides tracking coverage with high
efficiency for $|\eta| < 1$~\cite{coordinate_system}. Electromagnetic~\cite{CEM}
and hadronic~\cite{CHA} calorimeters surround the tracking system. They are
segmented in a projective tower geometry and measure the energies of charged and
neutral particles in the central region ($|\eta| < 1.1$).
A plug tile calorimeter covers the forward region ($1.1 < |\eta| < 3.6$).  
Each calorimeter has an electromagnetic shower profile detector positioned at
the shower maximum. The calorimeters are surrounded by muon drift
chambers~\cite{CMU}. The central muon detectors comprise four layers of
drift chambers which cover the region $|\eta|\leq0.6$. Forward muons, with $0.6
< |\eta| < 1.0$, are detected by an additional four layers of drift 
chambers. Gas \v{C}erenkov counters~\cite{CLC} measure the average number
of inelastic $p\overline{p}$ collisions per beam--crossing and thereby determine
the beam luminosity.  

We select $\ttbar$ candidates in data corresponding to
an integrated luminosity of $4.8\ \invfb$ in the lepton+jets
channel~\cite{ttbar_cross_section} by requiring one isolated
charged lepton --- an electron (reconstructed using the central
electromagnetic calorimeter) or a muon (reconstructed with the central or
forward muon detectors).

Primary leptons must have rapidity $|\eta|<1$. Electrons must have transverse
energy $E_\mathrm{T}>20\ \gev$ and muons must have transverse momentum
$p_\mathrm{T}>20\ \gev\!/c$. Four or more central jets ($|\eta| <2$)  with
$E_\mathrm{T}> 20\ \gev$ are also required. One of these jets must have a
secondary vertex displaced from the primary event vertex such that it is
``tagged'' by the SECVTX algorithm~\cite{secvtx} as being consistent with the
decay of a long--lived $b$-hadron. Jets with $b$-tags are restricted to $|\eta|
< 1$. Large missing transverse energy~\cite{met} is also required, $\met > 20\
\gev$. Jets corrected~\cite{jet_corrections} for multiple $p\overline{p}$
interactions in the event, non-uniformities in the calorimeter response along
$\eta$ and any non-linearity and energy loss in the uninstrumented regions of
the calorimeters. The $\met$ is corrected for the primary vertex location and
for any high--$p_\mathrm{T}$ muons in the event.

The contribution of non--$\ttbar$ backgrounds satisfying this selection has
been derived in precision measurements of the $\ttbar$ production cross
section~\cite{ttbar_cross_section}, these are tabulated in
Table~\ref{tab:background}. Also shown is the expected SM $\ttbar$ content,
given the event selection. However, in this analysis the SM $\ttbar$ content 
is estimated as the difference between the number of observed events and
the estimate of resonant $\ttbar$ events plus non--$\ttbar$ background
events. Table~\ref{tab:ndata} indicates the number of events observed 
in each of the subsamples considered in this analysis.  

\begin{table}[t]
  \newcommand\T{\rule{0pt}{2.6ex}}
  \newcommand\B{\rule[-1.2ex]{0pt}{0pt}}

  \centering
  \begin{tabular}{l|l|l}
    \hline
    \hline
    component        & $4$ jets           & $\geq 5$ jets      \\
    \hline
    non-$W$          & $ 46.1 \pm 35.7 $  & $ 15.7 \pm 12.2 $  \\
    $Z$+light flavor & $ 6.4 \pm 0.5 $    & $ 1.6 \pm 0.1 $    \\
    $W$+light flavor & $ 32.9 \pm 8.5 $   & $ 7.4 \pm 3.1 $    \\
    $Wb\overline{b}$ & $ 51.5 \pm 12.6 $  & $ 12.4 \pm 3.7 $   \\
    $Wc\overline{c}$ & $ 27.7 \pm 6.6 $   & $ 7.3 \pm 2.1 $    \\
    $Wcj$            & $ 14.0 \pm 3.3 $   & $ 3.0 \pm 0.9 $    \\
    single top       & $ 8.9 \pm 0.4 $    & $ 1.4 \pm 0.0 $    \\
    diboson          & $ 9.1 \pm 0.6 $    & $ 2.4 \pm 0.1 $    \\
    \hline
    total non--$\ttbar$ & $ 196.6 \pm 39.5$  & $51.2 \pm 13.3$ \\ 
    \hline
    SM $\ttbar$\T \B & $ 667.1 \pm 61.8 $ & $ 225.2 \pm 21.0 $ \\
    \hline
    \hline
  \end{tabular}
  \caption{Estimate of sample composition in terms of signal and background.}
  \label{tab:background}
\end{table}

\begin{table}[t]
  \centering
  \begin{tabular}{l|c|c|c}
    \hline
    \hline
                                 & central $e^{\pm}$ & central $\mu^{\pm}$ &
                                 forward $\mu^{\pm}$ \\ 
    \hline
    4 jets, 1 tag                & 480 & 221  & 131 \\
    4 jets, $\geq 2$ tags        & 110 & 33   & 21  \\
    $\geq 5$ jets, $1$ tag       & 164 & 81   & 41  \\
    $\geq 5$ jets, $\geq 2$ tags & 47  & 21   & 16  \\    
    \hline
    \hline
\end{tabular}
\caption{Number of events observed in each subsample for data corresponding to
  an integrated luminosity of $4.8\ \invfb$.}
\label{tab:ndata}
\end{table}

For each candidate event, we apply the $\ttbar$ hypothesis -- the
observed event kinematics are mapped to the parton level using the information
available in the SM QCD ``Matrix Element'' for $\ttbar$ production and
decay~\cite{PhysRevLett.100.231801}. 
By constraining the event kinematics 
to the hypothetical matrix element, we can reconstruct with enhanced precision
any kinematic quantity which can be expressed in terms of the parton momenta. 
In this analysis, we reconstruct the $\ttbar$ invariant mass probability density
function (pdf) for each event.
 
The probability for $\ttbar$ production and decay, given $p$,
the four--momenta of the decay products of the $\ttbar$ system, is 
\begin{equation}
  \label{eq:parton-level_probability}
  \pi(p|m_{t})
  =\frac{1}{\sigma{(m_{t})}}\int \mathrm{d}z_a \mathrm{d}z_b f_k(z_a) f_l(z_b)
  \mathrm{d}\sigma_{kl}(p|m_{t},z_a,z_b),
\end{equation}
where $f_k(z_a)$ and $f_l(z_b)$ are the parton density functions for
incoming gluons or quarks with flavor $k$ and $l$, with $z$--components of
parton momenta, $z_a$ and $z_b$. 
$\mathrm{d}\sigma_{kl}(p|m_t,z_a,z_b)$ is the differential cross section -- it is
proportional to the squared SM matrix element for $\ttbar$ production and decay.
In  this analysis, we fix the top-quark mass, $m_t = 172.5~\gev\!/c^2$. 

Any quantity which can be expressed as a function of the momenta of the top
decay products can be estimated as a probability density.  
We consider the invariant mass of the top--antitop pair, and derive the
following pdf of $x$, the reconstructed  $\ttbar$ invariant mass:   
\begin{equation} \label{eq:ttbar_pdf}
  \rho(x) \equiv \int \sum_{k} \pi(p_k|m_{t})
  W(j |p_k)\delta(x-\mathbbm{M}(p_k))\mathrm{d}p_k.
\end{equation}
$\mathbbm{M}(p_k)$ is the invariant mass of the top-antitop pair, given
parton configuration $p_k$, and $W(j |p_k)$ is a transfer function
which maps the observed jets $j$ to the parton level. The sum is over the number
of permutations, or jet--parton assignments, possible given the observed
event. 

The transfer functions $W(j |p_k)$ used in Eq.~(\ref{eq:ttbar_pdf}) are
estimates of the primary parton energy corresponding to the observed jet
$E_\mathrm{T}$. These estimates are derived from Monte Carlo simulation, where
jets are matched to partons within $\Delta R \equiv \sqrt{\Delta\eta +
  \Delta\phi} < 0.15$. No other jets or partons may be within $\Delta R < 0.6$
of the jet/parton match. We separate the response function into $10\ \gev$
jet--energy bins and into five bins of jet pseudo-rapidity. 

The matrix-element calculation produces a
probability density function of the reconstructed $\ttbar$ 
invariant mass for each event.
We use this sampled pdf as the observable in a likelihood calculation.
The probability for an event in sample $i$ is a function of the signal
fraction $f_{\mathrm{sig}}$ in the sample: 
\begin{equation}
  \label{eq:per-event_probability}
  P_i(f_{\mathrm{sig}}) = f_{\mathrm{sig}}P_{\mathrm{sig},i} + (1-f_{\mathrm{sig}})P_{\mathrm{bg},i}.
\end{equation}
Here we regard the hypothetical $Z'\goesto\ttbar$ component as signal and SM
$\ttbar$ and non--$\ttbar$ processes as background. 
The cross section $\sigma$ for $Z'\goesto\ttbar$ and the signal fraction are
proportional, 
\begin{equation}
  \label{eq:cross-section_signal_fraction}
  f_{\mathrm{sig}} = \frac{1}{n}\sigma \sum_{i=1}^{N_s}L_{i}A_{i}
\end{equation}
where $n$ is the total number of events observed,
$N_s=3\times2\times2=12$ is
the number of samples~\cite{number_of_samples}, and
$L_i$ and $A_i$ are, respectively, the integrated 
luminosity and absolute signal acceptance.
The signal and background probabilities $P_{\mathrm{sig},i}$ and
$P_{\mathrm{bg},i}$ are, according to their templates,
\begin{equation}
  \label{eq:P_sig}
  P_{\mathrm{sig},i} = \int T_{\mathrm{sig},i}(x)\rho(x)\mathrm{d}x,
\end{equation} and
\begin{equation}
  \label{eq:P_bg}
  P_{\mathrm{bg},i} = \int T_{\mathrm{bg},i}(x)\rho(x)\mathrm{d}x.
\end{equation}
These templates, $T_{\mathrm{sig},i}$ and $T_{\mathrm{bg},i}$, are the summed
event--by--event pdfs from each of the model components (e.g. the Monte Carlo
models of SM $\ttbar$, $W+$jets, etc.). The templates and the per--event pdfs
are normalized to unit area: 
\begin{equation}
  \label{eq:template}
  T_{c,i}(x) = \frac{1}{N_c}\sum_{j=1}^{N_c} \rho_j(x)
\end{equation}
\begin{equation}
  \label{eq:per-event_pdf_norm}
  \int\rho(x)dx \equiv 1
\end{equation}
where $T_{c,i}$ is the template for component $c$ in sample $i$ and $N_c$ is the
total number of events used to model component $c$. 

\begin{figure}[t]
 \includegraphics[width=\columnwidth]{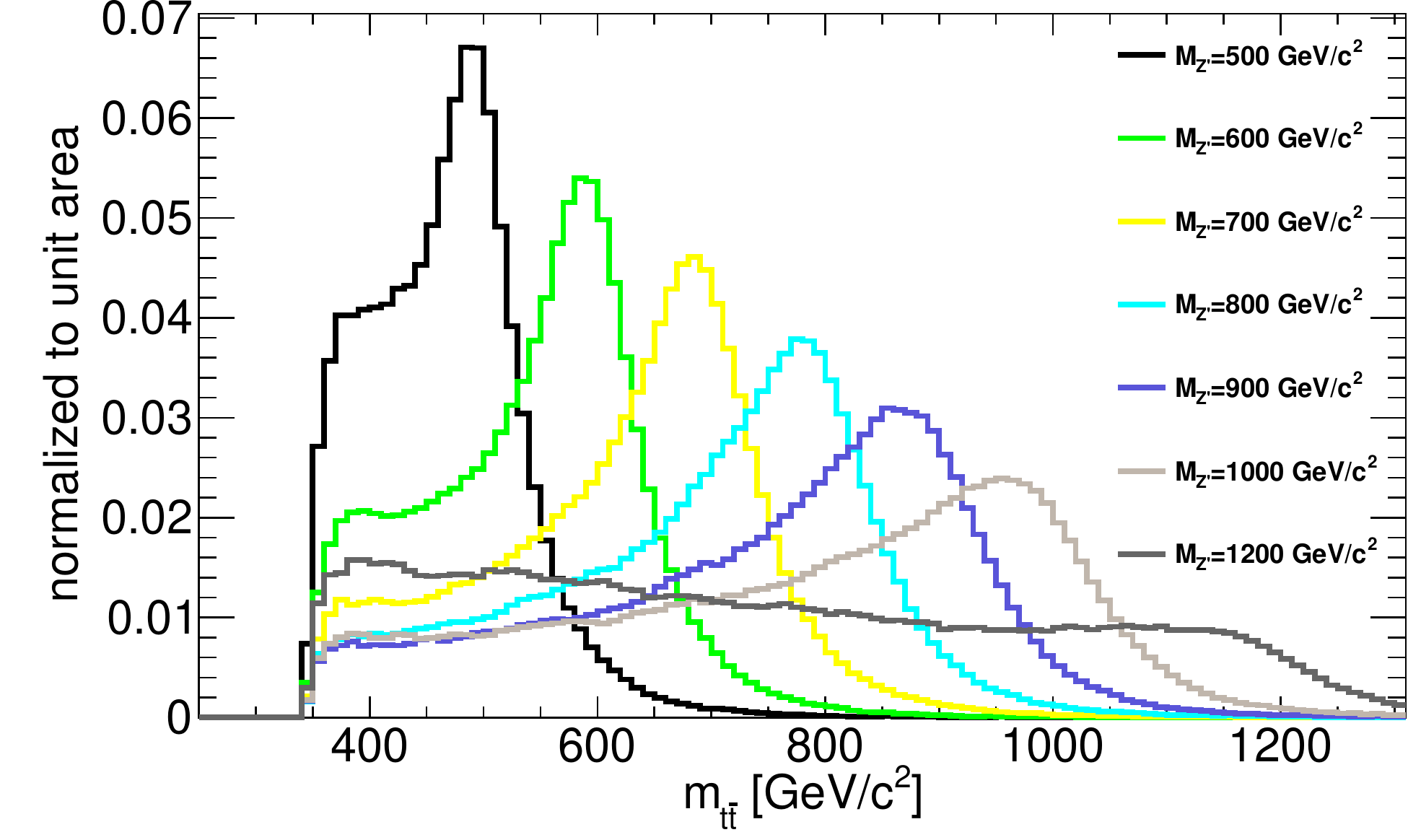} %
 \caption{Signal templates---summed over $b$-tag multiplicity, jet multiplicity 
          and primary lepton type.}
 \label{fig:templates}
\end{figure}

Templates for each model component are constructed by summing the per--event
probability densities. In Fig.~\ref{fig:templates} we show the templates for
$Z'\goesto \ttbar$ at several values of the $Z'$ pole mass, reconstructed
according to the procedure described above. Note that the mass peak is generally
prominent. For the samples where the $Z'$ pole mass is $1000\ \gev\!/c^2$ and
above, fewer $Z'$ particles are produced on--shell. The off--shell $Z'$ is
produced with mass according to the energy available from the parton density
functions, but decays to a real $\ttbar$ pair. The invariant mass spectrum for
on and off--shell $Z'\goesto\ttbar$ depends on the $Z'$ pole mass; we are able
to distinguish the $Z'$ mass, given the observed reconstructed $\ttbar$ mass
spectrum, even in cases where the $Z'$ mass is very large and the $Z'$ is
largely produced off--shell. 

To model  the background components 
{\sc alpgen}~\cite{Alpgen} version 2.10 with {\sc pythia} version
6.2.16~\cite{PYTHIA} parton showering is used for $W+$heavy flavor and
$W+$jets. 
The shape of the $\ttbar$ invariant mass at next to leading order (NLO) is
softer than the predictions of leading order Monte Carlo codes. 
To correct for this effect, {\sc pythia} weighted at the generator level by {\sc
  mcfm}~\cite{MCFM} (version 5.8) is used to model the SM $\ttbar$ component.   
The QCD ``fake'' component (where a jet is mis--reconstructed as an electron) is
modeled in data, where jets are selected which are electron--like in their
identification variables. 
Compared to the signal templates (Fig.~\ref{fig:templates}), we notice excellent
separation between background and signal shapes. 

We write the likelihood simply as the product of the per--event probabilities,
\begin{equation}
  \label{eq:likelihood}
   L(\sigma)=\prod_{i,j}P_i(\sigma)\cdot G(\nu_j|\overline{\nu}_j,\sigma_j)
\end{equation}
where $G$ is a Gaussian distribution, $\nu_j$ is nuisance parameter $j$ with
expectation $\overline{\nu}_j$ and uncertainty $\sigma_j$. 
We include among the nuisance parameters each lepton category acceptance and its
error, as well as the lepton trigger efficiencies. SECVTX tagging acceptance is
modeled by applying a scale factor to account for differences between data and
simulation. This scale factor appears as another nuisance parameter. We
integrate over the nuisance parameters to incorporate these systematic effects
into the likelihood.

The measurement is affected by uncertainties inherent to our model of the
background: the amount of initial-- and final--state radiation (ISR and FSR) and
the uncertainty on the jet energy scale~\cite{jet_corrections} of 
the calorimeter contribute most significantly, while systematic uncertainties
due to parton distribution function and color--reconnection
effects~\cite{Skands_Wicke_CR_0, Skands_Wicke_CR_1} are negligible. While the
value of the top quark mass determines the value at which the $\ttbar$ invariant
mass spectrum rises~\cite{Frederix:2007gi}, the shape of the \emph{tail} of the
$m_{\ttbar}$ spectrum is insensitive to variation of the top mass within its
uncertainty. This analysis is sensitive specifically to variations of the
tail of the $m_{\ttbar}$ spectrum---uncertainty on the value of the top quark
mass is not a significant source of systematic error. 

We treat systematic uncertainties due to acceptance effects as nuisance
parameters (cf. Eq.~(\ref{eq:likelihood})), while we convolve the likelihood
function to include systematic effects due to background model uncertainties. 

\begin{table}[t]
  \centering
  \begin{tabular}[c]{c|c}
    \hline
    \hline
    $Z'$ pole\\ mass~$[\gev\!/c^2]$ & $\Delta\sigma~[\mathrm{pb}]$ \\
    \hline 
    450  & 0.210\\
    500  & 0.183\\
    550  & 0.155\\
    600  & 0.153\\
    650  & 0.087\\
    700  & 0.058\\
    750  & 0.044\\
    800  & 0.030\\
    850  & 0.025\\
    900  & 0.020\\
    950  & 0.012\\
    1000 & 0.014\\
    1100 & 0.016\\
    1200 & 0.011\\
    1300 & 0.029\\
    1400 & 0.036\\
    1500 & 0.065\\
    \hline
    \hline
  \end{tabular}
  \caption{Total background model systematic uncertainties at each of the mass
    points considered. These values appear as the width of the Gaussian
    convolved with the likelihood in Eq.~(\ref{eq:convolved_likelihood}).}
  \label{tab:syst_total}
\end{table}

Background model systematic uncertainties are estimated by evaluating ensemble
tests where the experimental circumstances are reproduced and the
parameter in question (e.g. initial-- or final--state radiation, estimates of
background shape or composition) is varied within its error.  
The resulting distribution of the maximum likelihood estimate of signal
cross section form the basis for these estimates of systematic uncertainty.
We take the difference in the means of the $+1\sigma$ and $-1\sigma$
distributions as the error due to the underlying uncertainty at each mass value
of the hypothetical resonances considered. 
The final tabulation of estimated systematic uncertainties for this analysis is
given in Table~\ref{tab:syst_total}.

These errors are assumed to be Gaussian in nature. The likelihood of
Eq.~(\ref{eq:likelihood}) is convolved with a Gaussian distribution whose width
is equal to the quadrature sum of the individual systematic uncertainties:
\begin{equation}
  \label{eq:convolved_likelihood}
  L'(\sigma)=\int_0^{\infty} G(\sigma,\sigma',\Delta\sigma)L(\sigma')d\sigma'.
\end{equation}

\begin{figure}[t]
 \includegraphics[width=\columnwidth]{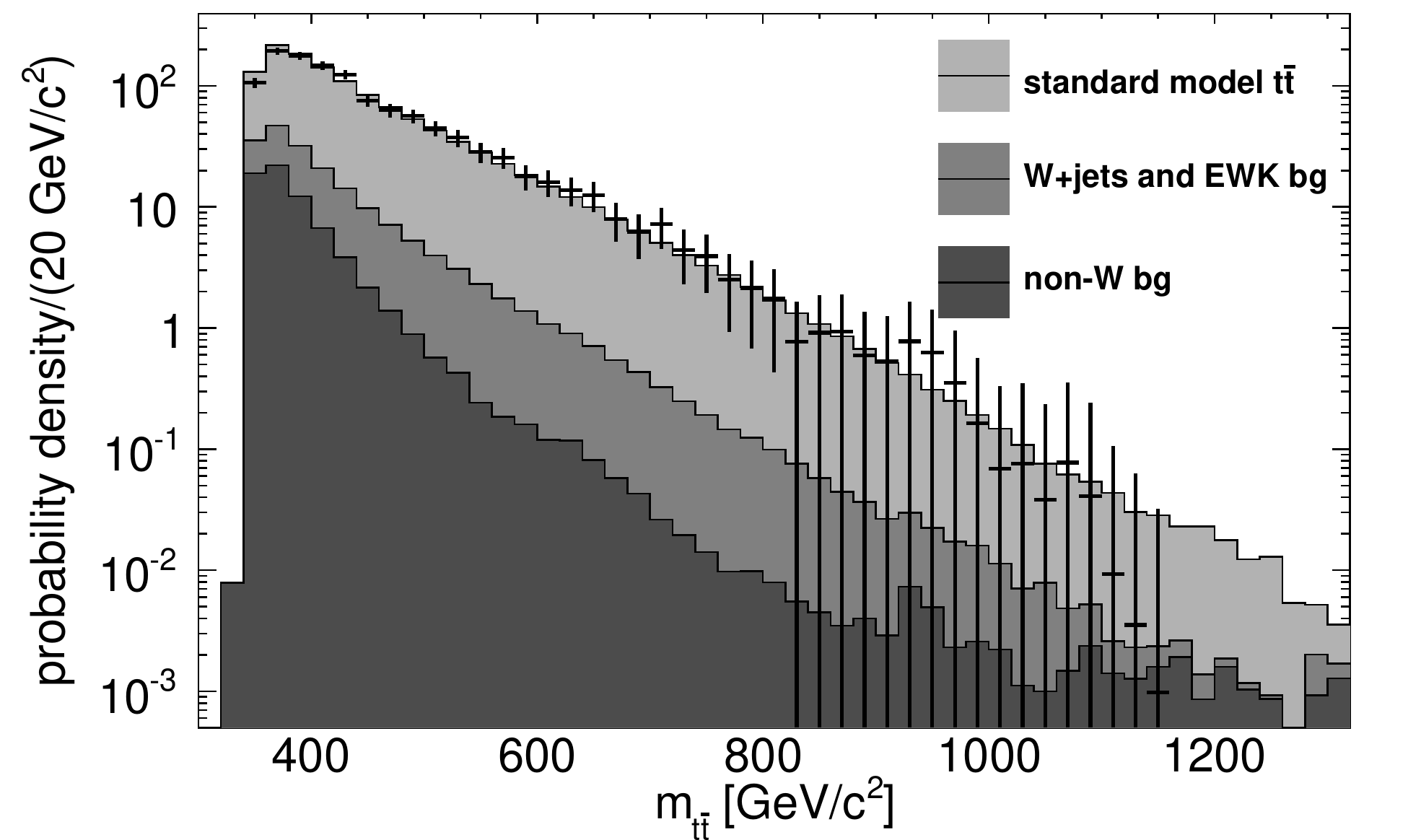}
 \caption{The histogram of total probability density for the 1366 $\ttbar$
          candidate events observed in $\mathcal{L} = 4.8\ \invfb$.}
 \label{fig:cData}
\end{figure}

\begin{figure}[t]
 \includegraphics[width=\columnwidth]{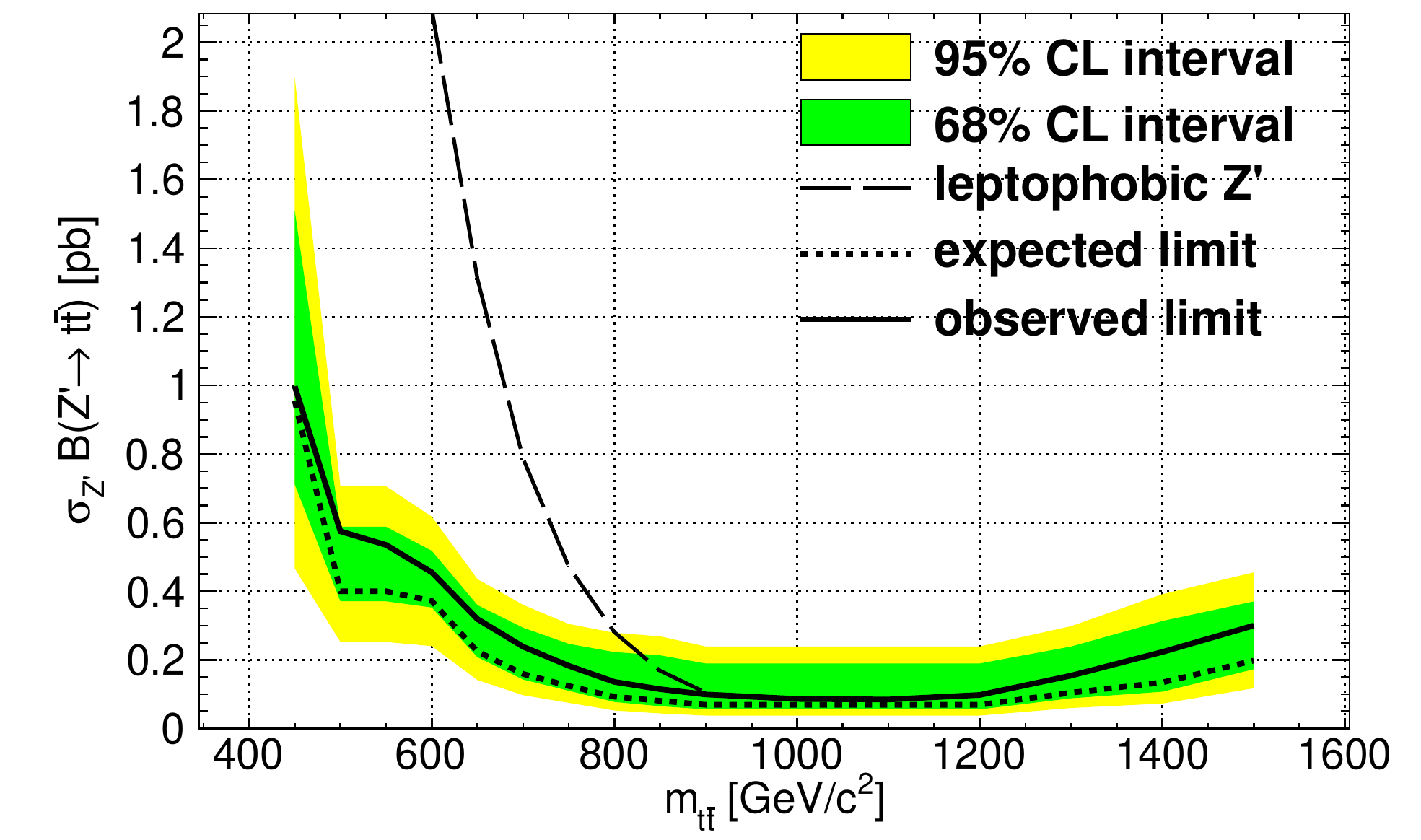}
 \caption{Expected and observed 95\% C.L. upper limit for
          $\sigma(p\overline{p}\goesto Z') \times BR(Z'\goesto\ttbar)$ for
          $\mathcal{L}=4.8\ \invfb$ of integrated luminosity as a function of
          reconstructed $\ttbar$ invariant mass. The solid line indicates the
          observed limit. The dashed line indicates the theoretical cross
          section for a leptophobic $Z'$~\cite{topcolor_Zprime}.}
 \label{fig:cLimits}
\end{figure}

The distribution of reconstructed $\ttbar$ invariant mass observed in the data
is shown in Figure~\ref{fig:cData}. This histogram shows the integrated
probability density in each $\ttbar$ invariant mass bin. For each event we
observe a {\em distribution}; as a consequence any single event can contribute
probability density to more than one bin. The data show no indication of
resonant $\ttbar$ production. Moreover, the distribution observed agrees very
well with the background model over five orders of magnitude.

At each value of the $Z'$ pole mass considered we calculate the Bayesian 95\%
confidence level (C.L.) upper limit on the value of the cross section times
branching ratio for $Z'$ production and decay to $\ttbar$. This upper limit is
the value of cross section times branching ratio which covers 95\% of the area
under the convolved likelihood, integrating from zero cross section, 
where a flat prior is assumed.  

\begin{table}[t]
  \centering
  \begin{tabular}[c]{c|c|c}
    \hline
    \hline
    $m_{Z'}\ [\gev\!/c^2]$ & Expected limit [$\pb$] & Observed limit [$\pb$]\\
    \hline
    450  &  0.954 & 1.05\\
    500  &  0.400 & 0.49\\
    550  &  0.400 & 0.46\\
    600  &  0.371 & 0.40\\
    650  &  0.224 & 0.26\\
    700  &  0.159 & 0.20\\
    750  &  0.123 & 0.15\\
    800  &  0.092 & 0.11\\
    850  &  0.080 & 0.10\\
    900  &  0.068 & 0.09\\
    950  &  0.069 & 0.08\\
    1000 &  0.069 & 0.07\\
    1100 &  0.069 & 0.07\\
    1200 &  0.070 & 0.08\\
    1300 &  0.104 & 0.13\\
    1400 &  0.134 & 0.18\\
    1500 &  0.197 & 0.24\\
    \hline
    \hline
  \end{tabular}
  \caption{Expected and observed 95\% C.L. limits on the cross section times
    branching ratio for narrow resonance production.}
  \label{tab:upper_limits}
\end{table}

Figure~\ref{fig:cLimits} shows the expected and observed limits on the $Z'$
cross section times branching ratio at each mass point considered. These data
are also given in tabular form in Table~\ref{tab:upper_limits}. The green
band shows the $\pm1\sigma$ expectation for the null hypothesis, the yellow band
the $\pm2\sigma$ expectation. Also included is a curve indicating the
theoretical cross section for a leptophobic $Z'$~\cite{topcolor_Zprime}. 

We have searched for narrow resonant states decaying to top--antitop pairs in a
sample corresponding to an integrated luminosity of $4.8\ \invfb$ of CDF II
data. Events were selected in the lepton + $\geq 4$ jets topology with at least
one jet tagged as coming from a $b$ quark. Monte Carlo samples were used to
model the appearance of narrow resonant states decaying to $\ttbar$. A
matrix--element reconstruction method was applied, and for each event a
probability density function of the reconstructed $\ttbar$ invariant mass was
sampled. This formed the basis for a likelihood fit to extract the cross section
times branching ratio for narrow resonance production, where non--$\ttbar$
background fractions are constrained according to their best estimates, and SM
and resonant $\ttbar$ components may vary according to the data. Systematic
uncertainties such as those arising from the error on the jet energy scale, or
uncertainties on parton distribution functions were incorporated by convolving
the likelihood function with a Gaussian with width equal to the
estimated uncertainty on the cross section times branching ratio due to the
underlying uncertainty. Systematic uncertainties arising from acceptance affects
such as trigger efficiencies or scale factors are treated as nuisance parameters
in the likelihood function. The benchmark model of a leptophobic
$Z'$~\cite{topcolor_Zprime} is ruled out at 95\% C.L. for $Z'$ masses below
$900\ \gev\!/c^2$. 
 
We thank the Fermilab staff and the technical staffs of the participating
institutions for their vital contributions. This work was supported by the
U.S. Department of Energy and National Science Foundation; the Italian Istituto
Nazionale di Fisica Nucleare; the Ministry of Education, Culture, Sports,
Science and Technology of Japan; the Natural Sciences and Engineering Research
Council of Canada; the National Science Council of the Republic of China; the
Swiss National Science Foundation; the A.P. Sloan Foundation; the
Bundesministerium f\"ur Bildung und Forschung, Germany; the Korean World Class
University Program, the National Research Foundation of Korea; the Science and
Technology Facilities Council and the Royal Society, UK; the Institut National
de Physique Nucleaire et Physique des Particules/CNRS; the Russian Foundation
for Basic Research; the Ministerio de Ciencia e Innovaci\'{o}n, and Programa
Consolider-Ingenio 2010, Spain; the Slovak R\&D Agency; the Academy of Finland;
and the Australian Research Council (ARC).%


\begin{thebibliography}{99.}
\bibitem{lepton_jets_note}
    This is the so-called ``lepton+jets'' topology. We
  consider leptonic decays including only final--state electrons and muons.
\bibitem{coordinate_system}
  We use a cylindrical coordinate system with its origin in the center
  of the detector and the downstream proton beam as reference axis, where
  $\theta$ and $\phi$ are the polar and azimuthal angles, 
  respectively, and pseudo-rapidity is $\eta = -\ln{\tan{\theta/2}}$. 
  The transverse energy $E_\mathrm{T}$ (momentum $p_\mathrm{T}$) is defined as
  $E\sin{\theta}$ ($p \sin{ \theta}$).
\bibitem{met} The missing transverse energy $\vec{\not\!\!{E_{T}}}$ is
  calculated as the negative vector sum of the energy in each calorimeter tower
  multiplied by a unit vector in the azimuthal direction of the
  vector sum. $\met$ is defined as the magnitude of
  $\vec{\not\!\!{E_{T}}}$.
\bibitem{number_of_samples}{The twelve samples are the product of three
  primary  lepton types (central electrons, central muons, and forward muons),
  two $b$--tag multiplicities (1 tag, $\geq 2$ tags), and two jet multiplicities
  (4 jets, $\geq 5$ jets).}
\end{thebibliography}
\end{document}